%% file: sample-sigconf.tex
\documentclass[sigconf, 10pt]{acmart}

\usepackage{booktabs} 
\usepackage{flushend}
\usepackage{array}

\setcopyright{rightsretained}

\begin{document}
\title{Long Range Communication on Batteryless Devices} 

\author{Simeon Babatunde}
\affiliation{%
  \institution{Clemson University}
  \streetaddress{P.O. Box 1212}
  \city{Clemson}
  \state{SC}
  \postcode{43017-6221}
}
\email{sbabatu@g.clemson.edu}

\author{Nirnay Jain}
\orcid{1234-5678-9012}
\affiliation{%
  \institution{Clemson University}
  \streetaddress{P.O. Box 1212}
  \city{Clemson}
  \state{SC}
  \postcode{43017-6221}
}
\email{nirnayj@g.clemsone.edu}

\author{Vishwas Powar}
\affiliation{%
  \institution{Clemson University}
  \streetaddress{1 Th{\o}rv{\"a}ld Circle}
  \city{Clemson}
  \state{SC}}
\email{vpowar@g.clemson.edu}

\renewcommand{\shortauthors}{}

\begin{abstract}
Bulk of the existing Wireless Sensor Network (WSN) nodes are usually battery powered, stationary and mostly designed for short distance communication, with little to no consideration for constrained devices that operate solely on harvested energy. On many occasions, batteries and beefy super-capacitors are used to power these WSN, but these systems are prone to service-life degradation and current-leakages. Most of the systems implementing super capacitors  do not account for leakages after exceeding the charge cycle threshold. Frequent battery maintenance and replacement at scale is non-trivial, labor-intensive and challenging task, especially on sensing nodes deployed in extreme harsh environments with limited human intervention. In this paper, we present the technique for achieving Kilometer range communication on batteryless constraint devices by harnessing the capabilities of LoRa technology.

The evaluation results show that long range communication with LoRa radio achieves an average of 7\% more Packet Delivery Rate (PDR) compared to CC1101 radio in a non Line-of-Sight (LoS) settings with constant transmit power of 5dBm. This work is one of the first to explore the possibilities of long range communication on batteryless devices that use tiny ceramic capacitors for energy storage. We conclude this paper by stating the challenges faced and recommendations in implementing LoRa based communication on batteryless devices. 

\end{abstract}

%



\maketitle

\input{samplebody-conf}

\bibliographystyle{ACM-Reference-Format}
\bibliography{sample-bibliography}

\end{document}

%% file: samplebody-conf.tex
\section{Introduction}
Majority of existing wireless sensor nodes deployed for applications like environmental monitoring \cite{OTHMAN20121204} are battery powered and mostly designed for short distance communication. As we scale to large deployments, the manual replacement of batteries becomes a bottleneck. Battery replacements can be mitigated using energy harvesting techniques that may vary based on the environment of deployed sensor nodes. Sometimes, as a replacement to batteries, sensing nodes utilize beefy super-capacitors which are prone to current leakages after exceeding the charge cycle threshold \cite{7356481}. To ensure a balance between the harvested energy and operational energy for a sensing node, intelligent adaptive algorithms for power management should be employed. Modular analysis of many such sensing systems clearly indicate that the most power consumption is directly associated the communication peripheral(either a radio or GPS). In this paper, we present the technique for achieving Kilometer range communication on constrained devices that are powered by harvested energy (Solar) for long term operation. The reason behind this effort is due to the fact that batteryless devices that gather energy from the environment and execute opportunistically usually promise a sustainable, maintenance-free, and environmentally-friendly alternative \cite{flickr}. However, these benefits don't usually come without some challenges. Furthermore, current state-of-the-art sensors for air quality monitoring are both costly and power hungry.  

In order to evaluate the efficiency and performance of our system, we design and implement a custom PCB prototype for solar energy harvesting and management. The board interface seamlessly with the MSP430FR5969 launchpad by Texas Instrument. The federated energy storage-UFoP \cite{Hester3} approach was adopted in designing the board. Each peripheral (LoRa radio and Air quality sensor) has a dedicated UFoP which allows automatic management of peripheral charging. We evaluated the system using air quality monitoring as our motivating applications. We carry out series of experiments to evaluate the impact of transmit power on other characteristics like transmission range, current consumption, data rate and line-of-sight (LOS) transmission. Also, a comparative analysis was carried out between the performance of SX1276 based RFM96W LoRa module and the popular CC1101 radio transceiver. 

We were able to accomplish over 1KM transmission range in a  pair-to-pair communication mode using the RFM96W LoRa radios on 433MHz. The Arduino Diecimila act as the TX node and Arduino Uno as the RX node. The arduino platforms were used due the challenges faced while trying to get MSP430 to work. The outcome of the comparison between LoRa and CC1101 shows that LoRa communication achieves better Packet Delivery Ration (PDR) and transmission range in a non-LOS setup with barriers and interference. We concluded by stating the challenges faced and recommendations on achieving long range communication on batteryless devices.

\noindent
\textbf{Contributions: } We explored the possibilities of implementing long range communication on batteryless devices. The research effort was directed towards understanding and harnessing the long range communication capability of the LoRa radio module. We make the following contributions in this paper:
\begin{enumerate}
  \item Presents a novel approach of enabling Kilometer(km) range communication on constraint devices powered solely by harvested energy (Solar).
  \item A custom hardware prototype design and implementation for energy harvesting and management. This prototype is a breakout board that powers the system for long range communication (LoRa) and it also interfaces seamlessly with the MSP430FR5969 launchpad by TI.
  \item Comprehensive evaluation and juxtaposition of LoRa performance with conventional CC1101 radio transceiver. This comparison cover metrics like transmit power, current consumption, Packet Delivery Ratio (PDR) and transmission range.
\end{enumerate}

\section{Related Works}
Existing research efforts have considered the possibility of enabling remote battery-less sensing using commodity devices like \textbf{Bluetooth Low Energy (BLE)} has dominated the space of mobile wireless communication \cite{ble}, especially for devices that
require low power operation. But it is not flexible enough for networks that suffer from intermittent power failures and also lacks the capability to transmit huge amount of data in a single packet \cite{ble2}.
\textbf{WiFi} is the technology of choice for wireless communication spectrum \cite{wiki}. Mostly used on high end devices with considerable amount of energy. Although WiFi supports high throughput and bandwidth \cite{wiki}, but a power consumption of 100mW at 20dB transmit power makes it unsuitable for use on low power devices which manage to scramble micro Amps of current. 
\textbf{ZigBee} is popular because of its lightweight, low-cost, low-speed, low-power protocol \cite{omojokun}. It can support up to 128bytes of data and a data rate of 250Kbps operated in the 2.4GHz band \cite{omojokun}. However, it requires development of algorithms for autonomous meshing of beacons \cite{joshi}, and also provides limited access to the physical layer properties. Also, Zigbee communication is limited to few 100 meters.

Most WiFi, Bluetooth, Zigbee and other wireless chipsets use 2.4GHz, which is great for high speed transfers \cite{adafruit}, but not suitable for low power long range communication. \textbf{LoRa technology} eliminates this challenge by enabling long range, low-power, low-bit rate wireless communication on end devices with limited energy \cite{aloy}. The novelty of our work is inherent in the ability to enable kilometer range communication on end devices that operate solely on harvested energy. Our work brings the best of long range communication and energy harvesting for enabling new applications that existing systems couldn't support. Our approach focus specifically on using LoRa radio (SX1276) for the long range communication on 433MHz licence-free frequency. Our motivating application uses the P2P LoRa communication mode. This enable us to configure one node as a transmitter (TX) and the other as a receiver (RX).

\section{Background}
Wireless Sensor and Actuator Networks (WSANs) plays a vital role of a inseparable technology of the Internet of Things (IoT), WSANs have emerged as a variation of Wireless Sensor Networks (WSNs). As an add on to the tasks performed by WSNs, WSANs are capable of keeping an eye on physical phenomenons (performing measurements e.g. vibration, humidity, radiation, article concentration levels,...), processing data collected by sensors, making decisions based on the sensed data and taking actions respectively (in our case, send data to remotely located base station).
For WSANs, IEEE 802.15.4 standard is the most widely-used communication scheme, which provides both physical
layer and Media Access Control (MAC) layer specifications (Figure 1).
This standard facilitates low-cost and low power transceivers,
but have a disadvantage over range of transmission that is only a few tens of meters. In
past few years, some wireless technologies enabling Long Range (LR) communication of several kilometers with power usage similar to WSAN nodes transceivers \cite{Martinez} have emerged. One of those technologies in use is LoRa \cite{semtech}, by the LoRa Alliance(Semtech). \textbf{LoRa (short for long range)} is a spread spectrum modulation technique derived from chirp spread spectrum (CSS) technology. LoRa operates in the
868/915 MHz ISM bands, promises great communication distances: upto 2 km in urban and
around 16 km in rural areas, and a bit-rate in the range between 0.37 and 46.9 kbps \cite{goursaud:hal-01231221}.

One of the main goals of this project is to acquire an easy, low power solution, interconnecting nodes with the base station setup within long distance \cite{augustine} and in order to meet these requisites LoRa technology seems like the best solution.
The technology offers a mix of long range, low power consumption and secure data transmission and
is gaining significant traction in IoT networks being deployed by wireless network operators. For our project purpose as we focus more on comparative study of the LoRa technology and CC1101 widely used long range, low power transceiver, we have made our system to work on frequency level of 433MHz. Since, CC1101 operates on this frequency, it'll be meaningful to evaluate both the transceivers on this frequency.

\subsection{Communicating with LoRa}

\begin{figure}[H]
\centering
\includegraphics[scale=0.80]{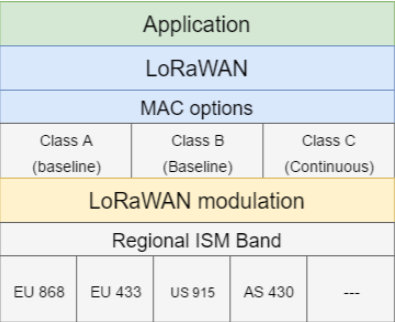}
\caption{LoRaWAN Stack layers}
\label{frame}
\end{figure}

LoRa technology is divided into two parts: LoRa physical layer and LoRaWAN. LoRa physical layer enables the physical layer communication whereas LoRaWAN is a low power, wide area networking protocol based on LoRa technology. It was designed to wirelessly connect battery based or remotely placed systems to the internet. The LoRaWAN takes advantage of the radio spectrum in Industrial, scientific \& medical ISM band. LoRaWAN also accounts for the power consumption of the remotely located node, the capacity of the network, quality of service and security.

Spread spectrum communication technique was opted to achieve better communication distance ranges, this spread spectrum technique uses wideband linear frequency modulated pulses whose frequency keeps changing over a specific time interval to encode information. Opting this technology has two benefits one of which being significant increase in receiver sensitivity due to the processing gain of the spread spectrum technique and the other being high tolerance to frequency misalignment between node and the base station.

\subsection{Network Architecture}
All remotely deployed systems want to transmit data as far as they can and to achieve this most of them uses mesh network architecture. In a mesh network, every specific node forwards information received from some other node to increase the communication range. But, this approach adds complexity to the network, reduces network capacity and also adds up to power consumption as all the nodes have to stay awake all the time to receive and transmit data, that too some nodes' information might be irrelevant for that node.

\begin{figure}[H]
\centering
\includegraphics[scale=0.39]{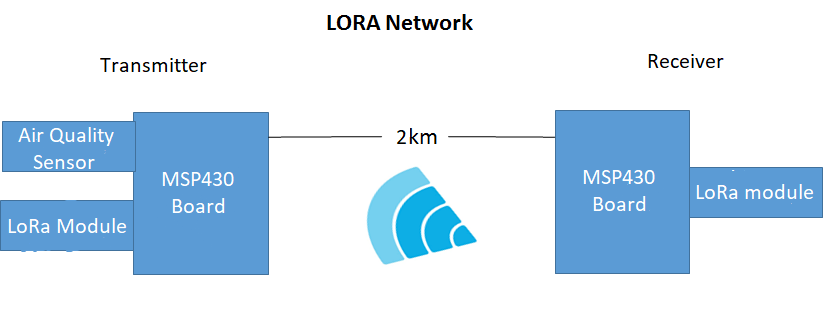}
\caption{LoRa P2P layout}
\label{frame}
\end{figure}

A node in LoRa on the other hand is individually connected to one or more gateways and sends information directly whenever needed and can go in sleep mode when doesn't have anything to do. Further, gateways forward information to the cloud server. Server gets redundant data and then it's job is to get rid of that redundancy. This mechanism results in power consumption \& increased network capacity. 
In our project we opted star topology and have created a test bed with one base station and one end node as shown in figure 2, to test P2P communication. Our system works on 433MHz instead of LoRa's proposed spectrum which is 868/915 MHz (depending upon the region of operation) so as to get justifiable comparison with P2P communication between two CC1101 transceivers.   

\subsection{Power Management}
Powering up the device deployed at remote locations has been a big challenge always, specially when the power supply is intermittent \citet{Hester2}. Researchers have come up with solutions for energy harvesting \cite{Hester} \cite{Sharma} this work will need something that can take decision on boost or not to boost the power harvested, which will be made to relate with the characteristics of the signals transmitted. Also, research and experiments have directed us towards deciding that use of multiple capacitors specifically dedicated for a peripheral unit\textbf{(UFoP - the United Federation of Peripherals)} is much better then deploying one shared capacitor\cite{Hester3}. 

\section{System Design}
This section covers the various design decisions we made in order to accomplish a system for adaptive long range communication on batteryless things while leveraging the unique features of LoRa module. Some of our design decisions were influenced by previous research efforts in the areas of batteryless sensing and Wireless Sensor Networks (WSN). Majority of existing research efforts either implement long range communication on a system with a dedicated battery or tetherd power. Others use super capacitors in collaboration with rechargeable batteries. We are the first to power a LoRa based long range sensing and communication systems with harvested energy on tiny ceramic capacitors.     

Our system provide solutions to some of the major challenges encountered in remote environmental monitoring applications like \textbf{outdoor air quality monitoring and algae bloom monitoring in fresh and marine water} \cite{airquality}; especially in constrained environments like space or arctic when human intervention is highly limited. Despite the challenges attributed to the intermittent nature of energy harvesting \cite{Hester}, our system appears to be robust enough to handle variations in energy availability through dynamic adaptation.

The system design is made up of three main sections: Energy Harvesting, Computing Core and Peripherals. Since the MCU, LoRa radio and the air quality sensor can operate between 1.8V and 3.6V, we designed the energy energy harvesting section to supply 3.3v to all peripherals as well as the MCU. Our initial design was based on 3V level appears insufficient for the RFM96W LoRa breakout board. The breakout board already has a 3.3V regulator, thus making it challenging to power the radio with 3V. The figure below shows simplified block diagram of the system architecture.

\begin{figure}[h!]
\centering
\includegraphics[scale=0.40]{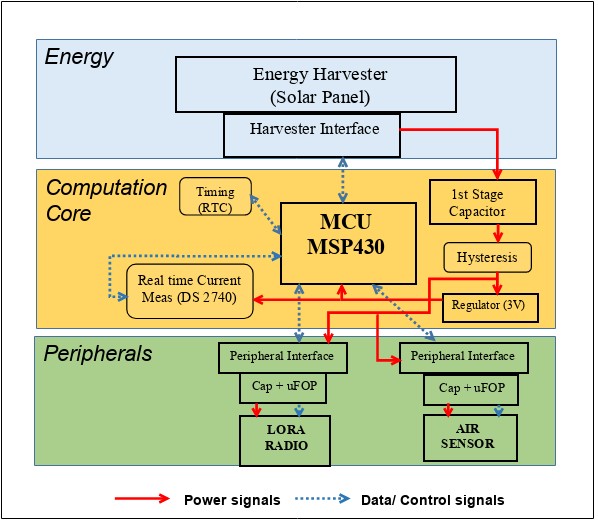}
\caption{System Block Diagram}
\label{frame}
\end{figure}
The Energy section comprises of the energy harvesting circuitry which supplies power to whole system. It interfaces the solar panel output with the federated energy storage (UFoP) of the peripherals and the micro-controller unit. The computation section comprises of the MCU, timing circuit as well as the power sensing circuitry. The final stage (peripherals) consist of the UFoP circuitry for both the LoRa radio and the air quality sensor. 

\section{Implementation}
In order to evaluate the performance and viability of long range communication on batteryless devices, we designed a prototype sensor node with long range communication and energy harvesting capabilities. The prototype was designed as a breakout board which can be seamlessly interfaced with MSP430FR5969 launchpad by Texas Instrument. We used air-quality monitoring as the motivating application for this work. The system design consist of three major stages, which are energy harvesting, air-quality measurement, and long-range communication which is based on LoRa module. In this section we describe the main specifics of the hardware and software implementation of our prototype. 

\subsection{Energy Harvesting} This stage consist of components and modules needed for energy harvesting and management. The federated energy storage-UFoP \cite{Hester3}approach was adopted in designing the energy harvesting section. UFoP enables automatic management of peripheral charging  and also allows for dynamic retasking and reprioritization of energy resources at runtime \cite{flickr}. We assigned dedicated UFoP for each of the peripherals on our prototype (LoRa and Air quality sensor). Energy harvesting is done using sunboy new energy's solar panel rated as 6V 70mA. The solar panel supplies the main capacitor (100uf) which is protected by a 5.1V Zener diode. An hysterisis chip MIC841 was introduced after the main capacitor to ensure stable operation i.e. the system turns on above 3.38V and turns off below 3.05V. The systems was designed to operate at 3V level, so we used 3V S-1313 series voltage regulators.

\subsection{SX1276 LoRa Radio Module} This stage contains the UFoP designed to power the radio module. We want the radio module to start charging after all other peripherals have started charging. A combination of voltage divider circuit and a TLV3691 comparator was used to configure charging threshold for the radio UFoP as well as for other peripherals. The Adafruit RFM96W LoRa Radio Transceiver-433 MHz was used for the long range communication. The LoRa module was interfaced with the microcontroller via SPI. The radio UFoP contains an 100uF capacitor which is sufficient to hold substantial amount of charge needed to transmit radio packets. The interrupt pin goes high when the capacitor is fully charged so that the MCU can close the UFoP GATE in order to use the radio for transmission. The RFM96W LoRa breakout board which is based on SX1276 LoRa module exposes eight pins (MOSI, MISO, CS, CLK, RESET, DIO-0, VCC and GND) for coordination with the MCU. 

\begin{figure}[H]
\centering
\includegraphics[scale=0.4]{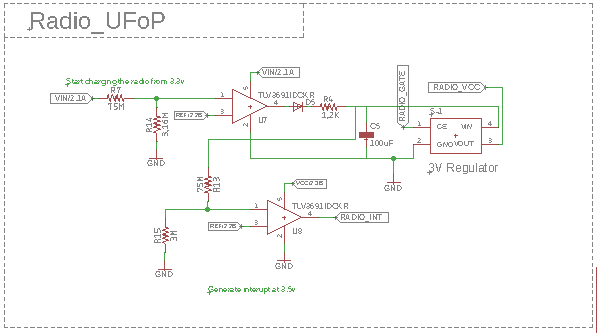}
\caption{Radio UFoP designed to charge the capacitive storage for the LoRa radio module from 3.3V and generate interrupt at 3.5V}
\label{frame}
\end{figure}

\subsection{Air-Quality Measurement} 
\begin{figure}[H]
\centering
\includegraphics[scale=0.6]{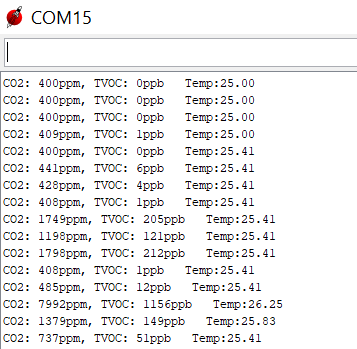}
\caption{Air Quality Sensor output}
\label{frame}
\end{figure}
This stage include the UFoP specifically designed for the air quality sensor. The UFoP initiates the charging of the 22uF capacitor when the input voltage gets to 3.2V and generates interrupt when it's fully charged. We used the Adafruit CCS811 Air Quality Sensor Breakout which can measure Volatile Organic Compounds (VOC) and equivalent Carbon dioxide (eCO2) via I2C. The interrupt pin goes high when the capacitor is fully charged and the MCU closes the gate whenever air quality measurement is to be taken. The CCS811 air quality sensor breakout board exposes some pins for I2C communication with the MCU and it is based on ams unique micro-hotplate technology which enables a highly reliable solution for gas sensors, very fast cycle times and a significant reduction in average power consumption (as low as 1.2mW) \cite{ams}. 

The prototype was designed as a breakout board that fits perfectly on the MSP430FR5969 launchpad. The rational behind this results from need to minizie the amount of components on the custom board by harnessing existing MCU on the launchpad . This also helped to eliminate the need for extensive use of jumper cables between the board and MCU. The MCU is powered externally from the MCU\_CC (3V) pin on the breakout board. This pin is connected to the VCC pin of the jumper J12 and the J10 jumper (Power Select) is switched from debugger to external. A 50 ohm SMA antenna connected to the LoRa radios ensures the radio packets can travel long distance. We used ISL21080 voltage reference to generate a 125mV signal which is compared with the output of the voltage dividers to keep state.   

\begin{figure}[H]
\centering
\includegraphics[scale=0.06]{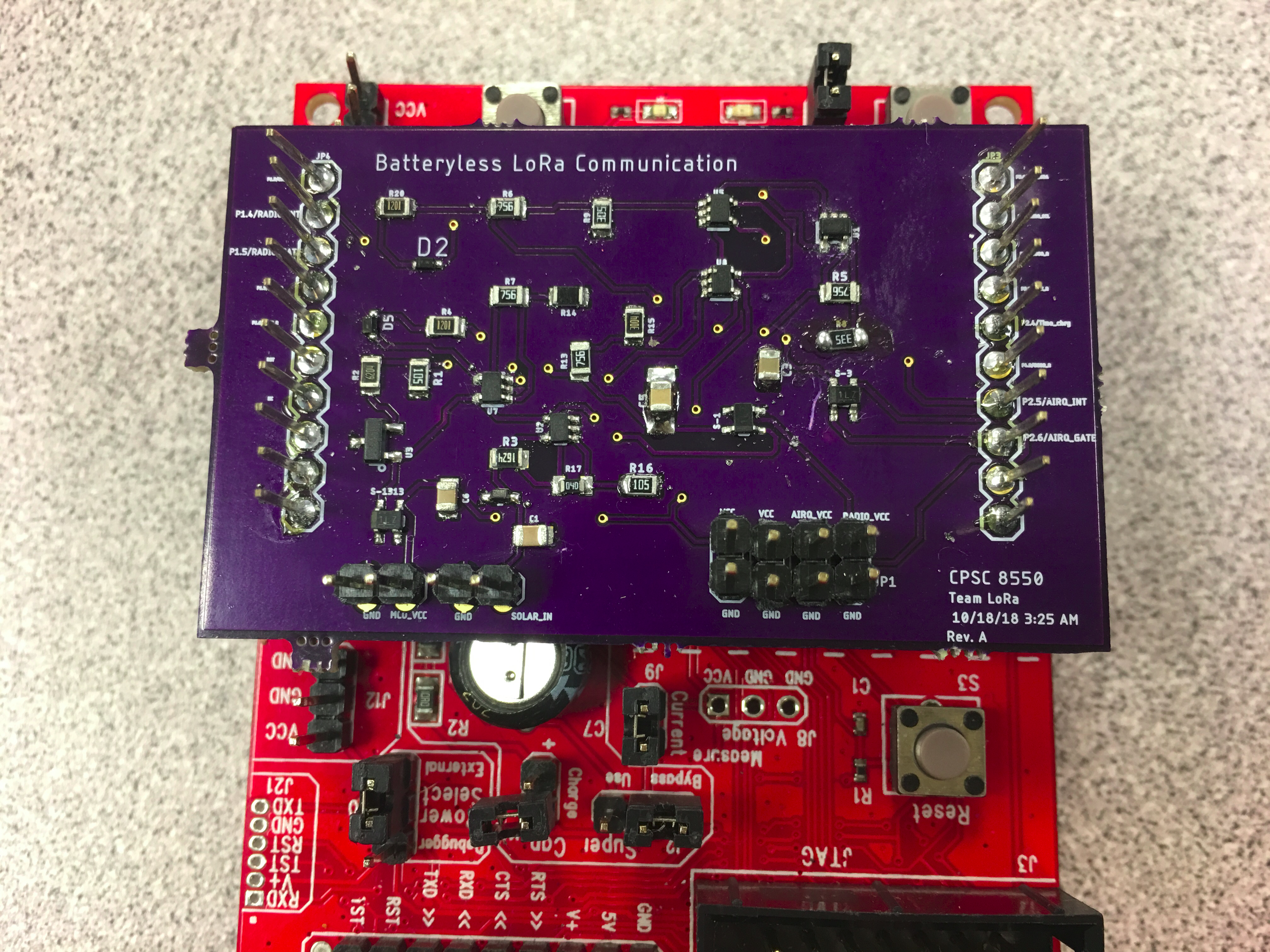}
\caption{The breakout board PCB interfaced with MSP430FR5969 launchpad.}
\label{frame}
\end{figure}

\subsection{Software} We utilized embedded C/C++ to implement all the necessary operations on the MCU. Both the Energia platform as well as the Arduino platform were used during code implementation. The MCU was programmed with the logic that coordinates the operation of the UFoP as well as air quality sensing and LoRa data transmission and receiving. Sensing and transmission takes place every 10 seconds. The MCU goes to sleep after each sensing and transmission cycle. Whenever the UFoP capacitors are fully charged, the signal on the designated interrupt pin goes high which allows the MCU to close the AIRQ\_GATE for air quality measurement and RADIO\_GATE for long range transmission. The LoRa module was configured to operate on 433MHz, this provides us with the opportunity to perform comparative analysis between the LoRa module and the CC1101 radio. 

We implement the prototype design on Autodesk Eagle 9.0.1 and generate gerbers files, which were later sent to OSHPARK for PCB printing. Majority of the components used conforms with the 0805 package size. 

\section{Experimentation}
Our project was all about testing an idea and to see if it can work and if it works how efficient it is in comparisons to the systems working out there on harvested energy. As communication among wireless sensor network is one of the areas where energy consumption is a very big issue specially when the system is powered with intermittent energy, we believe in finding a solution for it by testing our system in those real time scenarios. 

\begin{figure}[H]
\centering
\includegraphics[scale=0.08]{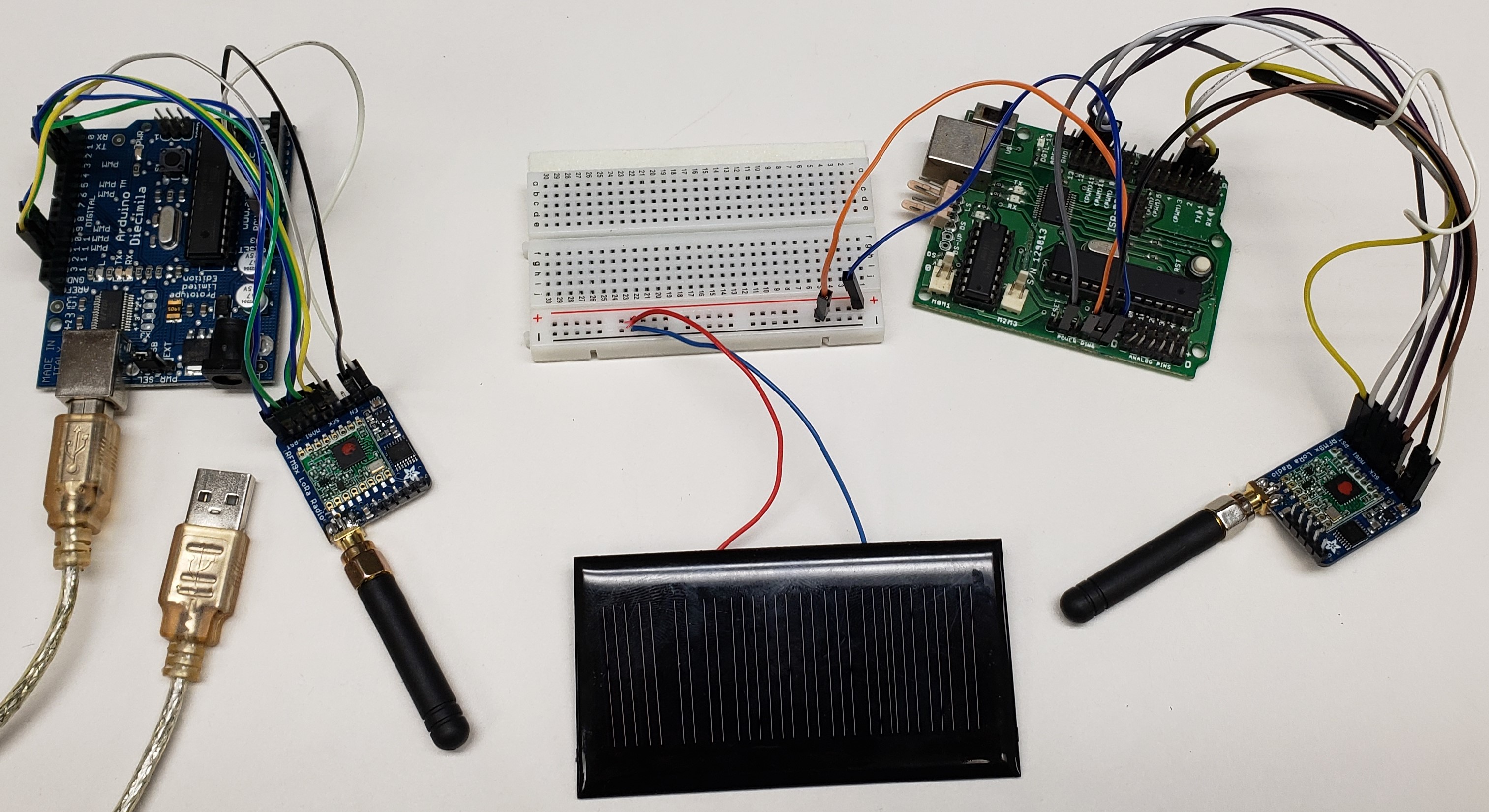}
\caption{LoRa Nodes}
\label{frame}
\end{figure}

\subsection{Current Consumption}
We started the comparative study of our system with transceiver CC1101 with this experiment, which was to measure the current consumed by it at different transmission power levels. All experiments were done for P2P communication. These measurements were taken by varying the transmission power of the LoRa transceiver in the software part and checking current consumed by the system for each of these configurations. In total 5 transmitting power configurations were tested and readings were noted, this experiment was done by keeping the receiver and transmitter at same distance throughout the experiment so that the readings can be genuine and factors like orientation of radio and line of sight doesn't effect the readings.  

\begin{figure}[H]
\centering
\includegraphics[scale=0.1]{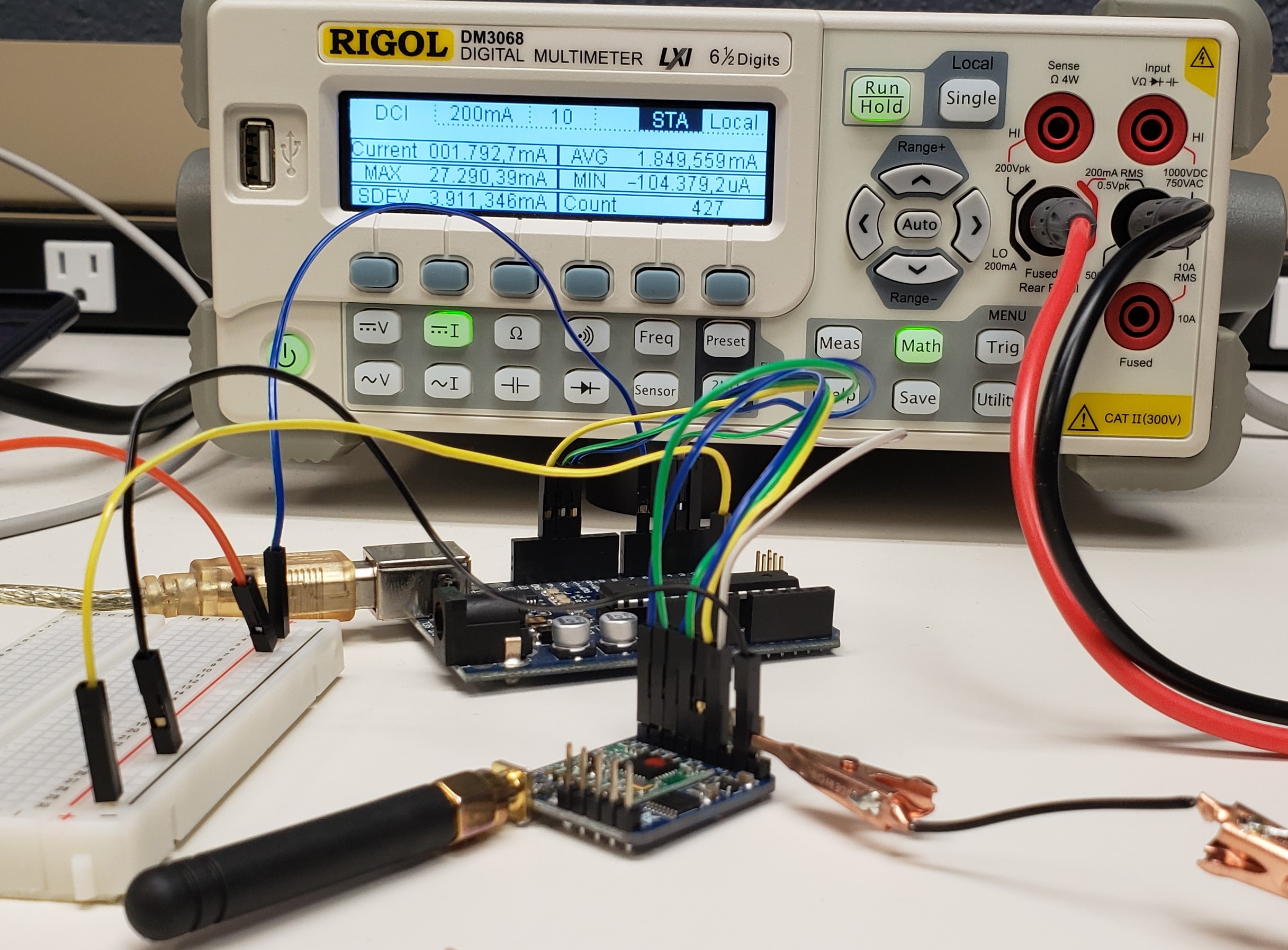}
\caption{Current Measurements with maximum, minimum and average currents for transmitter }
\label{frame}
\end{figure}

\subsection{Range and Transmission Power}
Our 2nd experiment was a vital one which gave us the idea about the range of LoRa transceivers with respect to transmission power of LoRa radios. This experiment was set up in two different settings first being the non \textbf{line of sight (LOS)} setting in which the transmitter was fixed at a specific location and the receiver was carried around a route which was kept same for all the transmission power configurations and the second being the line of sight setting where again transmitter was fixed at a specific location and receiver was taken away from the transmitter in a straight path making sure both the radios are visible to each other.

\subsubsection{Non-LOS Configuration}
This was a very wild test in which nothing like the orientation, height or position of the radios was considered and the receiver was just carried around a fixed route which was randomly chosen. Experiment was started with keeping the transmission power as 5dBm and the receiver was carried along the route, one by one multiple transmission powers in the interval of 5dBm (lowest limit of LoRa) - 23dBm (highest limit of LoRa) were chosen and the behaviour of the communication channel was observed. For each transmission power, the current consumption of the system was also monitored so as to get an idea if transmitter put some extra efforts to transfer the packet successfully when the receiver is taken a little far away from it.

\subsubsection{LOS Configuration}
This configuration was tested in a proper set up, where transmitter was fixed at a position and receiver was carried in a straight path away from the transmitter. Also, apart from the pre decided straight path some other factors like the orientation of the antenna, height of the antenna all were maneuvered so as to keep the radios in line of sight of each other. Similar to the prior configuration in this configuration also the transmission power was varied from 5dBm to 23dBm.

\subsection{Power Supply}
One of the important conclusion we wanted out of this project was if the LoRa is suitable for batteryless field or not. To test this we conducted two experiments, first one was with stable/continuous power supplies which were table top power unit and USB power supply from a laptop and the second was solar panel. We tried both to see how the results vary and what all problems can arrive.

\subsubsection{Continuous Power Supply}
Under this experiment we supplied power to our system using table top power supply and USB port of the laptop, both seemed to work well without any issue. Also, we limited the current to 16mA and our system worked fine for 5dBm of transmission power at which the maximum current consumption was around 15.6mA and average consumption was around 11mA when our microcontroller was running at 3.7V.

\subsubsection{Intermittent Power Supply}
For this test bed we tested multiple configurations of solar panels. We started with using 3 solar panels each of 6V 70mA in indoor setting, they worked but we had to flash light on them otherwise they were not harvesting enough to suffice the current requirement of the system. As soon as we took them outdoors in normal daylight they worked well and we were able to setup proper connection between both the LoRa nodes. Then the number of solar panels were decreased from three to two and we were still getting all the data packets properly. Ultimately, we tested with one solar panel but it didn't harvested enough energy to start the communication though the system was getting enough power to wake up but it got stuck in infinite booting loop. We were able to make our system work perfectly with single solar panel also but we had to put it directly under bright sunlight.

\subsection{UFoP Bank Capacitor}
Actual project was supposed to include MSP430FR5969 launchpad but we faced few issues with it because our launchpad was designed using UFoP approach. In UFoP approach each peripheral has it's own power supply unit and the regulator in it makes sure that the peripheral only gets the supply when there is enough energy stored in bank capacitor to power it up. 

The problem we faced was we made our UFoP to supply 3V to our LoRa radio(by putting the 3V regulator), when we connected the LoRa radio to the peripheral breakout board Vcc for radio the voltage dropped down to 1.9V which shouldn't have happened and this much voltage was not enough to power up the radio nodes. In efforts to make it work we narrowed down the problem to two factors that might have been responsible for the failure, either the bank capacitor of the LoRa UFoP or the regulator that we were using to provide 3V supply to LoRa radio. For solving the issue we started with using voltage booster but it itself consumed some current and reduced the voltage. Next we tried changing the bank capacitor to 100uF, 1000uF in parallel with the originally placed 100uF capacitor, none of them worked. Then we tried placing only 1000uF capacitor instead of 100uF, that also didn't give us results. One more thing we wanted to try was higher rating voltage regulator but it was not readily available in the time frame we had. So, we had to modify our system and use the LoRa radios with arduino instead, which somewhere accounted for a little more power consumption in our system as MSP430 boards work on relatively less power than arduino.

\begin{table}[h!]
\caption{List of components used}
\renewcommand{\arraystretch}{1.5}
\small\addtolength{\tabcolsep}{-2.85pt}
\begin{tabular}{|m{2cm}|m{1.9cm}|m{1.7cm}|m{1cm}|m{1cm}|}
	\hline
	\textbf{Module} & \textbf{Model} & \textbf{Availability} & \textbf{Cost(\$)} & \textbf{Count}  \\
	\hline
    Lora Radio & RFM96W & Adafruit & 19.95 & 2\\
	\hline
    Air Quality Sensor & CCS811 & Adafruit & 19.95 & 1\\
	\hline
	Voltage Regulator & SC82AB & Digikey  & 0.851 & 1\\
    \hline
    MCU & MSP430FR5969 & TI & 15.99 & 2 \\
    \hline
    Current Measurement & DS2470 & Mouser & 2.47 & 1 \\
    \hline
	Hysteresis Chip & MIC841 & Digikey & 0.58 & 1 \\
    \hline
	Zener Diode & SOD-523 & Digikey & 0.0.643 & 1 \\
    \hline
    IC Comparator & SOT 353 & Digikey & 1.01 & 3 \\
    \hline
    Solar Panel & 6V, 70mA & Amazon & 5.95 & 1\\
    \hline 
    
\end{tabular}
\end{table}


\section{Evaluation}
 We evaluate our system with ground truth results obtained from line-of-sight(LOS) measurements and through-wall(Non-LOS) settings for measuring parameters like range and current consumption for varying transmission powers.
\subsection{Current Consumption}

\begin{figure}[htp]
\centering
\includegraphics[width=8.7cm]{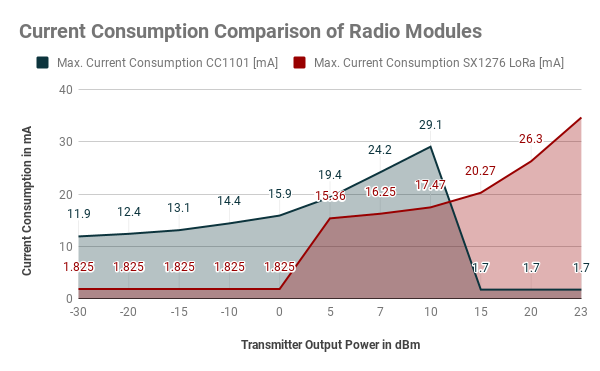}
\caption{Current Consumption(mA) Comparison of Radio Modules with respect to transmission power(dBm)}
\label{fig:lion}
\end{figure}

While measuring the current for receiver and transmitter it was clear that steady state current (Mode0: when it neither transmits nor receive and goes to sleep mode) for LoRa nodes is 1.8mA. Also, we kept an eye on the current trends, the way they change, it was pretty much clear that they follow the same trend of increasing current gradually and then reduction in the current consumption. This variation in the current consumption was independent of the transmission activity and also didn't depend on the distance between receiver and transmitter. We checked it by moving the receiver away from transmitter and placing it side by side, movement of receiver apparently doesn't matter.
As can be seen in Figure 8, the current consumption of LoRa radio is less than the CC1101 transceiver on same transmission powers.  

\subsection{Range and Transmitter Power}

\begin{figure}[htp]
\centering
\includegraphics[scale=0.42]{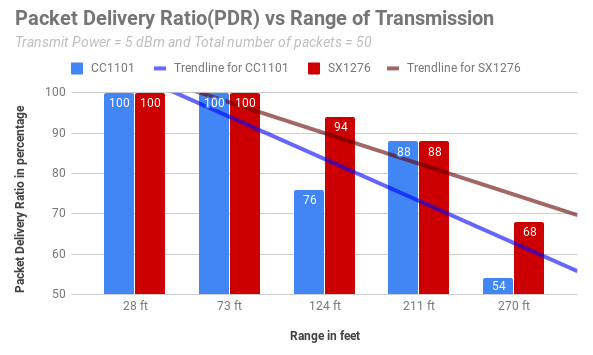}
\caption{Packet Delivery Ratio Comparison}
\label{frame}
\end{figure}
\subsubsection{Non-LOS Configuration}
Figure 10 shows the relationship between the PDR and transmission range on both the LoRa module and CC1101 radio. LoRa achieves 7\% more PDR in a Non-LOS settings while keeping the trasnmit power constant at 5dBm. This shows that LoRa can tolerate more interference than CC1101 even in the presence of obstacles like buildings. 
\subsubsection{LOS Configuration}
Figure 11 shows that based on the readings we obtained for CC1101 and for LoRa by our experiments CC1101 has a better range over LoRa at 433MHz. But, one thing to be noted is CC1101 starts working at -30dbM and only works till 10dBm whereas LoRa starts working at 5dBm and works till 23dBm. So, the transmission power which is the starting point for LoRa is one of the highest transmitting power of CC1101. 
 
Figure 12 clearly describes about the Range of LoRa transmission which we achieved at different transmitter power levels. 

\begin{figure}[htp]
\centering
\includegraphics[width=8.5cm]{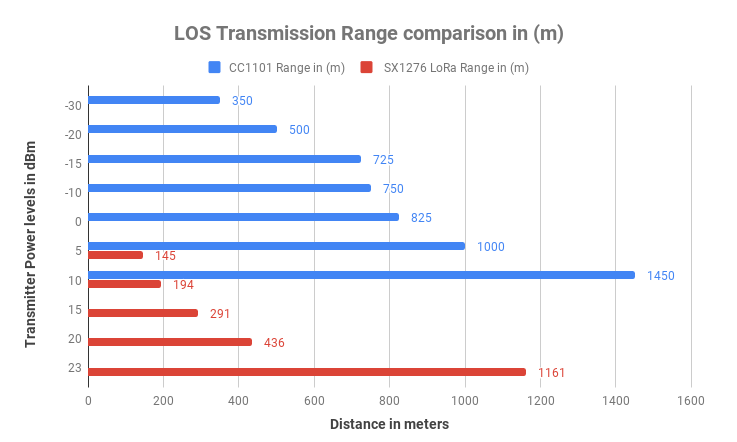}
\caption{LOS Transmission Range Comparison at specific transmission power levels}
\label{figure}
\end{figure}

\begin{figure}[htp]
\centering
\includegraphics[scale=0.44]{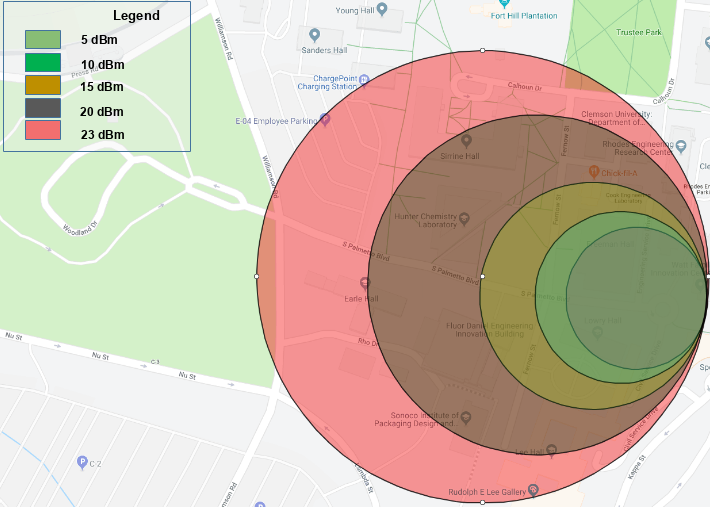}
\caption{Measured Approximate Range of LoS LoRa Communication}
\label{fig:lion}
\end{figure}

\subsection{Power Supply}
Continuous power supply doesn't posses any problem and the system works fine. Also, it's safety can be insured by limiting the current. Whereas, in case of USB power supply if somewhere in the system the connections are not proper and if by mistake there is a shorted line, it can damage the USB port.

While intermittent power supply is a new approach for these kind of systems as they may consume more power in comparison to other peripherals. Along with being new it's a tricky one too because if the system power is not managed properly the system might go into infinite booting loop and it may not be feasible to repair it if the node is placed somewhere remotely. Proper Adaptive algorithms need to be implemented on it so that the system doesn't stay awake for a very long time and just wake up whenever the interrupt is fired so as to conserve energy.

\subsection{Current consumption at different stages of LoRa protocol}
We tried to evaluate the current consumed by the LoRa transceiver for each stage of operation. For this approach we used a current sense resistor along with an Oscilloscope to measure the current waveform of the entire LoRa communication. The main motive behind this experiment was to analyze the behaviour of different LoRa stages with respect to current consumed. We used a sense resistor of 1.5 ohms for our experiments and observed an interesting characteristic of LoRa transceiver. Initially after we  powered our transmitter and set it to transit LoRa packets every 5 seconds, we noticed that without the presence of an active receiver on the channel, the transmitter idle current draw was significantly higher. We set-up the transmitter for a cold start each time it is powered on, i.e. the it undergoes the initialization process, handshaking modes, authentication and then transmission. This was done in attempt to bifurcate the current consumed in each of these processes. However, we couldn't successfully evaluate time required for each of these processes but could draw some useful conclusions. 
1. The transmitter would consume significantly higher mA currents when idle in "SEARCH MODE" than in an active transmission mode when discrete link between TX-RX is established.
2. Current consumption for receiving an ACK from the receiver was significantly lower than transmission currents.
We need accurate nsec - msec timing measurements to isolate each section of LoRa protocol to carry out in-depth power measurement analysis. Also, the LoS and non LoS distance between the transceiver and receiver might also play a huge role in packet time and handshaking signals.
\begin{figure}[h]
\centering
\includegraphics[scale=0.25]{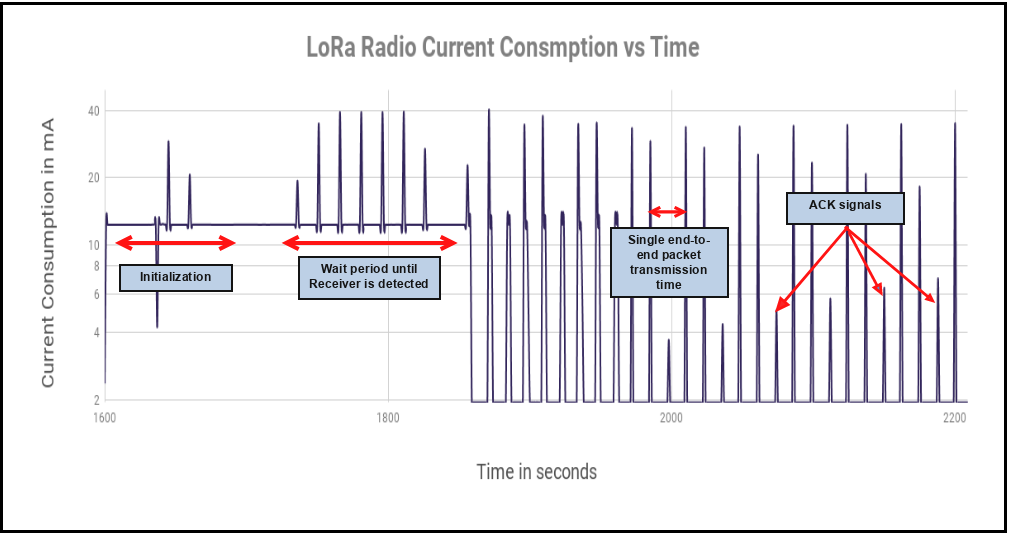}
\caption{Current consumed vs Time for LoRa Communication}
\label{fig:lion}
\end{figure}

\section{Discussion and Future Work}
\subsection{Challenges \& Recommendations}
\begin{enumerate}
  \item We faced issues while taking line of sight readings, as we had to keep the orientation of the radio pretty much same all the time otherwise we were losing connection in between receiver and transmitter in case of long ranges. 
  \item We couldn't find a long LOS distance to test LoRa Radio to it's limit. We believe it would have worked much better if the readings were taken in an large open field.  
  \item The 100uF capacitor used in the radio UFoP was unable to store enough charge to allow complete transmission of LoRa packet. We increased the values of the capacitor up to 1000uF but still unable to achieve full packet transmission.
  \item Our UFoP for LoRa radio didn't supply enough voltage so as to power it up. Three ways this could have been possible were; firstly the system was not getting enough current so voltage was just going down because of large current consumption just to maintain the power supplied, second reason could have been the bank capacitor was not getting enough time to charge back to full capacity and before it can charge to full capacity the radio drew whole energy so it was just staying at stable low voltage supply, third reason which came out to be the actual problem was that one of our component can't sustain the amount of current that our radio needs and it's just trimming it off. We realized it because our radio was getting initialized but not sending packet as shown in Figure 15 when we bypassed the voltage regulator and as soon as we bypassed the hysteresis chip (Figure 14) it started sending packets.
\begin{figure}[h]
\centering
\includegraphics[scale=0.20]{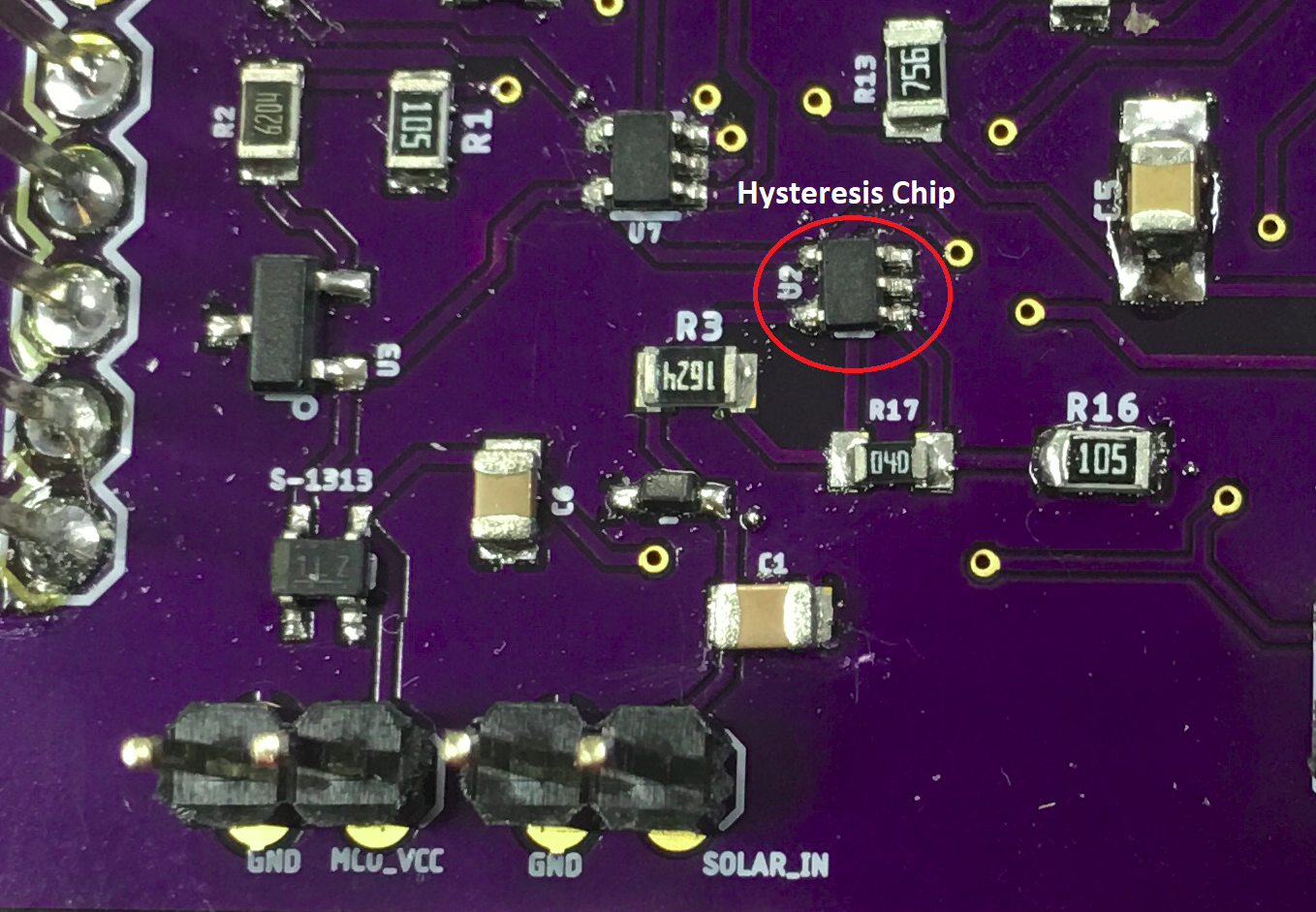}
\caption{Hysteresis Chip which was clipping current at 20mA}
\label{fig:lion}
\end{figure}

\begin{figure}[h]
\centering
\includegraphics[scale=0.40]{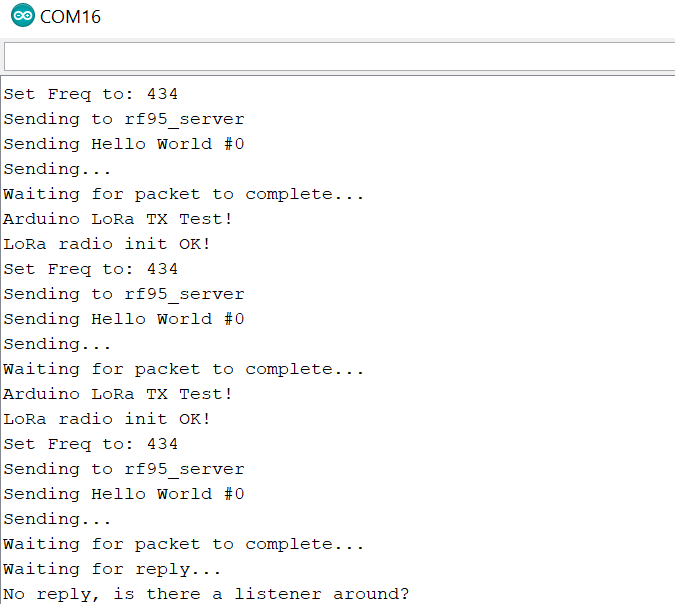}
\caption{LoRa Radio initializing but not sending packet because of lack of current}
\label{fig:lion}
\end{figure}  
  We troubleshooted the problem by checking each component one by one that we used on our UFoP, and our hysteresis chip could only allow 20mA whereas our radio needs more than that often. So, the voltage was getting pulled down because of lack of current supply. 
  
\end{enumerate}

\subsection{Future Work}

We wanted to have Stage 4 in our project which was focused at additional features for our project involving but not limited to multiple nodes communicating with LoRaWAN gateway, integrating water quality and quantity sensors, exploring the frequency domain of 868 MHz frequency band for potential LoRa communication and be able to perform adaptive data storage and transmission. These goals can be termed as ambitious goals of our project. The idea of implementing multiple sensor nodes communicating with a central gateway for achieving LoRaWAN was our initial focus for this project. But understanding the level of complexity involved in installing a LoRaWAN gateway and designing a communication channel for multiple sensor nodes under given time frame, we planned to achieve this functionality at a later stage. Knowing the power requirements involved for P2P communication of our LoRa radio and having a comparative analysis of energy requirements, we can then test the gateway to single node and then multiple node challenges. The gray area of working in 868 MHz frequency band is also what we would like to explore and compare results with European counterparts using the same frequency for LoRa communication. 

	Another fundamental functionality that we initially aimed at achieving was water quality and quantity measurements to monitor algae bloom problems. After a successful literature review of the available CO2 and dissolved O2 sensors, we found that majority of the sensor work on 3.3V-12V ranges requiring current draw of 100mA-2A range. These sensors are bulky and heavy and have very restricted interfacing options. A majority of sensors are power hungry and need SDI-12, MODBUS or 4-20mA communication standards. As the major niche of this project is aimed towards achieving LoRa long distance communication on federated energy storage, we have potentially planned on having air quality monitoring as a first check-point from application point of view. Further work and better circuit designs would be needed to meet water sensors energy demand and we plan on interfacing them at a later stage in the project.

\section{Conclusion}
Our project was aimed at accomplishing kilometer range long distance communication for batteryless devices using the P2P LoRa communication protocol. Based on the comparative tests, we found out that the SX1272 LoRa Radio provided better transmit range and better packet delivery rate(PDR) as compared to CC1101. But the transmit current for LoRa was significantly high and thus some further work is needed to get it to work efficiently on long distances. The area of LORAWAN is still to be exploited and studying behaviour of multiple nodes with a gateway would be a potential field of research. We would like to consider water quality measurement on power hungry sensors in order to evaluate our system under extreme conditions. A breakthrough in this area will enable new sets of interesting applications that current technologies cannot support.